\documentstyle[11pt]{article}
\begin{document}
\title{The Breakup Condition of Shearless KAM Curves\\ 
in the Quadratic Map}
\author{Susumu SHINOHARA\thanks{E-mail: susumu@aizawa.phys.waseda.ac.jp} 
~and Yoji AIZAWA\\
Department of Applied Physics, Waseda University, Tokyo 169}
\maketitle
\begin{abstract}
\indent
We determined the exact location of the shearless KAM curve in the
quadratic map and numerically investigate the breakup
thresholds of those curves in the entire parameter space.
The breakup diagram reveals many sharp singularities like fractals
on the reconnection thresholds of the twin-chains with rational
rotation numbers.
\end{abstract}
\section{Introduction}
\indent 
In this paper we focus on the complicated phenomena which occur in the
hamiltonian systems beyond the twist condition.  Because of the
violation of the twist condition\cite{rf:1,rf:2}, the KAM theorem and
the Poincar\'e-Birkhoff theorem do not hold near the shearless KAM
curve\cite{rf:8}, where the rotation number takes an extremum value.
When a perturbation is added, very complicated phase space structures
are observed in the vicinity of the shearless KAM curve: twin resonant
chains appear on both sides of the shearless KAM curve, and at a
certain parameter value the separatrices of the twin-chains merge.
These phenomena, so-called reconnections of twin-chains, are generic
ones in many nontwist systems\cite{rf:3}-\cite{rf:8}.\\
\indent
The quadratic map, which is the simplest example of a nontwist system,
has been studied from the viewpoint of the reconnection of
twin-chains, and the breakup mechanism of the shearless KAM curve has
been studied to a certain extent\cite{rf:3,rf:4,rf:8}.  However the
detailed phase space structures and the breakup condition of shearless
KAM curves in the entire parameter space have not yet been elucidated
completely.\\
\indent 
In this paper, carrying out with the quadratic map, we derive the
exact location of the shearless KAM curve and numerically determine
the breakup threshold of shearless KAM curves in the entire parameter
space.\\
\indent
The quadratic map is the following two-dimensional area-preserving map
$(I_n,\theta_n)$ $\mapsto$ $(I_{n+1},\theta_{n+1})$,
$$
T: \left\{
\begin{array}{ll}
I_{n+1} = I_{n} - K \sin(\theta_{n}),\\ \theta_{n+1} = \theta_n +
f_{\mu}(I_{n+1}), \hspace {6mm}(\mbox{mod} \hspace {2mm} 2 \pi)
\end{array} \right. \eqno(1) 
$$
$$f_{\mu}(I)=2 \pi \mu-I^2, \hspace{3cm}$$ 
where $K$ represents the strength of the perturbation, and $\mu$ is
the rotation number of the shearless KAM curve in the integrable
case,~i.e.,$K=0$.  Because the variable $\theta$ is $2\pi$-periodic,
$\mu$ is also periodic with period $1$.  Thus it is enough for us to
consider the parameter region $-0.5 \leq \mu < 0.5$.\\
\indent 
The quadratic map $T$ can be rewritten as $T=M_2 M_1$ by use of the
product of two involutions $M_1$ and $M_2$, where
$M_1^2=M_2^2=1$\cite{rf:4,rf:8} and,
$$
M_1: \left\{
\begin{array}{ll}
I^{\prime} = I - K \sin(\theta),\\ \theta^{\prime} = -\theta, \hspace
{6mm}(\mbox{mod} \hspace {2mm} 2\pi)
\end{array} \right. \eqno(2)
$$
$$
\hspace{2cm} M_2: \left\{
\begin{array}{ll}
I^{\prime} = I,\\ \theta^{\prime} = -\theta + 2\pi\mu - I^2. \hspace
{6mm}(\mbox{mod} \hspace {2mm} 2\pi)
\end{array} \right. \eqno(3)
$$
\noindent
Moreover the mapping $T$ has the symmetry $S$, i.e.,
$ST=TS$\cite{rf:8}, where
$$
S: \left\{
\begin{array}{ll}
I^{\prime} = -I,\\ \theta^{\prime} = \theta - \pi.
\hspace{6mm}(\mbox{mod} \hspace {2mm} 2\pi)
\end{array} \right. \eqno(4)
$$
\section{The reconnection of twin-chains}
\indent
Since the twist function $f_{\mu}(I)$ is quadratic, the mapping $T$
does not satisfy the twist condition ($d f_{\mu}/ d I$ $\ne$ 0 for
$\forall$ $I$), and the twin resonant chains appear on both sides of a
shearless KAM curve\cite{rf:3,rf:4,rf:8}.  In Figs.1 and 2, period-one
and period-two twin-chains are shown for various values of $K$, with
$\mu$ fixed.  When we change the value of the parameters $\mu$ or $K$,
the twin-chains exhibit topologically different structures; Fig.1(a)
for $K=0.05$ shows that two twin-chains with period-one are generated
on both sides of the shearless KAM curve, and Fig.1(c) for $K=0.20$
shows the saddle-node fusion of those two twin-chains.  At the
critical parameter value $K=0.1173906615$, the separatrices of these
twin-chains merge in the manner of the double homoclinicity(see
Fig.1(b)).  For general cases of odd periodic twin-chains, the merging
process occurs in the same way as described above.  The parameter
values at which the twin-chains merge are called reconnection
thresholds.  The reconnection thereshold for the period-one
twin-chains, $K_{rec}^{(1)}$, has been derived by Howard and Hohs by
using the averaged Hamiltonian method as follows\cite{rf:3,rf:4}:
$$ K_{rec}^{(1)} = \frac{2}{3} (2\pi\mu)^{3/2}. \eqno(5)$$
Although Eq.(5) was derived by the perturbation theory, the results
explain the numerical data very well\cite{rf:3}.\\
\indent
On the other hand, the twin-chains with even-periods reveal quite
different features from twin-chains with odd-periods.  Figure 2 is the
case for the merging of the twin-chains with period-two.  In the case
of the even periodic twin-chains, the pair of the resonant chains
appear with the same phase near the shearless KAM curve.  The
topological structure of the twin-chains is quite different from the
odd periodic case.  At a critical parameter value, the hyperbolic
periodic points of the twin-chains merge(see Fig.2(b)).  We will call
this state the reconnection for the twin-chains with even-periods.
The reconnection threshold for the period-two case, $K_{rec}^{(2)}$,
has been analytically derived as\cite{rf:4}
$$ K_{rec}^{(2)} = 2 \sqrt{2\pi(\mu+\frac{1}{2})}\quad. \eqno(6)$$
\section{The symmetry of the quadratic map and the shearless KAM curves}
\indent
Hereafter we focus our attention on the KAM curves between twin-chains
and derive the location of the shearless KAM curve.  Denote the KAM
curve which goes through the point $(\theta,I)=(0,I_{0})$ as ${\cal
K}_{I_{0}}$, i.e.,
$$ {\cal K}_{I_{0}} = \{ (\theta,I) | (\theta,I)=T^{n}{\bf x}_0, {\bf
x}_0=(\theta=0,I=I_0), - \infty \leq n \leq \infty \}. \eqno(7)$$
\indent 
The KAM curves are characterized by the rotation number $R$,
$$ R=R(T,(\theta_0,I_0))\stackrel{\rm def}{=}\lim_{N\to\infty}\frac1{2\pi
N}\sum_{n=1}^N(\theta_n-\theta_{n-1}), \eqno(8)$$
which is a function of the mapping $T$ and initial conditions
$(\theta_0,I_0)$. By fixing the value of $\theta_0$, the KAM curve is
parametrized by the value of $I_0$.\\
\indent
In the quadratic map, the KAM curve remains invariant under the
involutions $M_1,M_2$, which is shown as follows.  By using the
relation $M_{j} T^{n} = T^{-n} M_{j}(j=1,2)$ for any integer $n$,
$$ M_1 T^n {\bf x}_0 = T^{-n} M_1 {\bf x}_0 = T^{-n} {\bf x}_0
\eqno(9)$$ and
$$ M_2 T^n {\bf x}_0 = T^{-n} M_2 {\bf x}_0 = T^{-n}
(\theta=2\pi\mu-I_{0}^{2},I=I_0) = T^{-(n-1)} {\bf x}_0, \eqno(10)$$
where ${\bf x}_0=(\theta,I)=(0,I_0).$ Therefore we have
$$ 
M_{j} {\cal K}_{I_0} = \{ (\theta,I)|(\theta,I)=M_{j} T^n {\bf x}_0,
{\bf x}_0=(\theta=0,I=I_0), -\infty \leq n \leq \infty \} \eqno(11)
$$
$$ 
= \{(\theta,I)|(\theta,I)=T^n {\bf x}_0, {\bf x}_0=(\theta=0,I=I_0), -
 \infty \leq n \leq \infty \} \eqno(12)
$$
$$ = {\cal K}_{I_0}.\hspace{5mm} (j=1,2)\hspace{68.5mm}$$
\indent
The shearless KAM curve which passes through the point
$(\theta=0,I=I^{*}_{0})$, say ${\cal K}_{I^{*}_0}$, is defined as the
KAM curve whose rotation number $R^{*}$ satisfies the condition
$$R(T,(\theta=0,I=I_0)) \ne R^* \hspace{5mm} \mbox{for} \hspace{2mm}
\forall \hspace{0.5mm} I_0 \ne I^{*}_{0}.\eqno(13)$$ 
This definition is trivial in every part of the phase space if there
exists only one shearless KAM curve in the phase space.\\ \indent
Moreover, it is shown that the shearless KAM curve is invariant under
the transformation $S$ as follows.  By using the relation $S T^m = T^m
S$, we have
$$ R(T,S(\theta_0,I_0))=R(T,(\theta_0,I_0)). \eqno(14)$$
Therefore, if the shearless KAM curve goes through the point
$(\theta_0,I_0)$, it also goes through the point $S(\theta_0,I_0)$,
i.e., the shearless KAM curve is an invariant set of $S$:
$$S {\cal K}_{I^{*}_{0}} = {\cal K}_{I^{*}_{0}}. \eqno(15)$$
\indent
We require that the shearless KAM curve ${\cal K}_{I^*_0}$ crosses
$\theta$-axis at two points, denoted by $A:(\theta,I)=(\theta_{A},0)$
and $B:(\theta,I)=(\theta_{B},0)$ respectively.  Since the shearless
KAM curve is an invariant set of $M_2$ and $S$, the following two
equations hold:
$$ \theta_{B}=-\theta_{A}+2\pi\mu, \eqno(16)$$
$$ \theta_{B}=\theta_{A}-\pi. \eqno(17)$$ 
From the above equations, we have
$$
\theta_A=\pi(\mu+\frac{1}{2}), \eqno(18)
$$
$$
\theta_B=\pi(\mu-\frac{1}{2}), \eqno(19)
$$
which are $K$ independent.  Equations (18) and (19) determine the
exact location of the shearless KAM curves in the phase space.
\section{The breakup of the shearless KAM curves}
\indent
In order to obtain the breakup threshold of shearless KAM curves, we
numerically iterate the point $(\theta,I)$ $=$ $(\pi(\mu+1/2),0)$ for
each set of $\mu$ and $K$, and examine whether the motion is bounded
or not.  In cases when the motion is unbounded, the shearless KAM
curve is considered to have been destroyed.  In our numerical
calculations, the shearless KAM curve is considered to be broken up if
the absolute value of $I$ exceeds 2 during $10^5$ iterations.\\
\indent
The numerical results are shown in Fig.3, where the motion is bounded
in the black region, but the shearless KAM curve does not exist in the
white region.  This straightforward numerical method enables us to see
the breakup condition of the shearless KAM curves.  This method cannot
detect the detailed threshold boundary for the breakup of the
shearless KAM curves.  However, the result will be much more refined
by improving the breakup criterion in the numerical calculations.
Actually, the breakup thresholds obtained in Fig.3 are numerically
confirmed by the determination of the rotation number for the nearby
KAM curves.\\
\indent
The breakup threshold of the shearless KAM curve characterized by the
inverse golden mean $1/\gamma$ rotation number has already been
determined by del-Castillo-Negrete et al. using the Greene's residue
method\cite{rf:8}.  In the case of the critical value for the
destruction of the $1/\gamma$ shearless KAM curve, they obtained
$\mu_c=-0.313951$ and $K_c=1.54156$, which are indicated by the point
$(\mu_c,K_c)$ in Fig.3.  The shearless KAM curve for $\mu_c$ and $K_c$
is plotted in Fig.4, which is obtained by numerically iterating the
point $(\theta=\pi(\mu_c+1/2),I=0)$.\\
\indent
At the reconnection threshold of twin-chains, the rotation number of
the shearless KAM curve, which exists between two twin-chains, is
equal to the rotation number of the twin-chains.  Thus, the
reconnection threshold for the twin-chains with any period can be
obtained by numerically calculating the rotation number of shearless
KAM curves.  In the cases of period-one and period-two, it is possible
to derive the reconnection threshold analytically, as was shown
previously.  The $P/Q$ lines plotted in Fig.3 are numerically
determined reconnection thresholds for twin-chains with the rational
rotation numbers $P/Q$.  In the quadratic map, the rotation number of
a shearless KAM curve depends on the values of both $\mu$ and $K$.
This is because $K$ not only represents the strength of the
non-integrable perturbation\cite{rf:7}, but also the change in the
integrable part of the system.\\
\indent
As shown in Fig.3, the boundary of the breakup diagram reveals many
sharp singularities on the lines of the reconnection thresholds with
rational rotation numbers $P/Q$.  Figure 5 shows a magnification of
the breakup diagram around $(\mu_c,K_c)$, where fine singular
structures are seen.  We expect that much more fine structures,
corresponding to the reconnection thresholds for twin-chains with
higher periods, are immersed in the breakup diagram.
\section{Summary and discussion}
\indent 
In this paper, we have determined the exact location of the shearless
KAM curve in the quadratic map.  Furthermore, we succeeded in
obtaining the global structure of the breakup diagram of shearless KAM
curves in the entire parameter space.  The breakup diagram has many
sharp singularities, and they are well explained by the reconnection
thresholds for twin-chains.  Fine structures as shown in Fig.5 suggest
that the breakup diagram might be a fractal.  The theory which
combines the breakup of shearless KAM curves with the reconnection of
twin-chains remains to be established.\\
\indent
As mentioned by del-Castillo-Negrete et al., a shearless KAM curve
exhibits self-similar hierarchical structures for certain parameter
values\cite{rf:8}.  Now that the exact location of a shearless KAM
curve is obtained, it can be used to study the geometry of shearless
KAM curves.  The detailed analysis of the breakup diagram and the
geometrical properties of shearless KAM curves will be reported
elsewhere.
%  References

\vskip 1cm
\section*{Figure captions}
{\bf Fig.1.}
\begin{minipage}[t]{145mm}
The merging process of the twin-chains with period-one for various
values of $K$ at $\mu=0.05$.  (a)$~K$=0.05(below
threshold). ~(b)$~K$=0.1173906615(reconnection
threshold). ~(c)$~K$=0.20(above threshold).
\end{minipage}
\vskip 3mm
\noindent
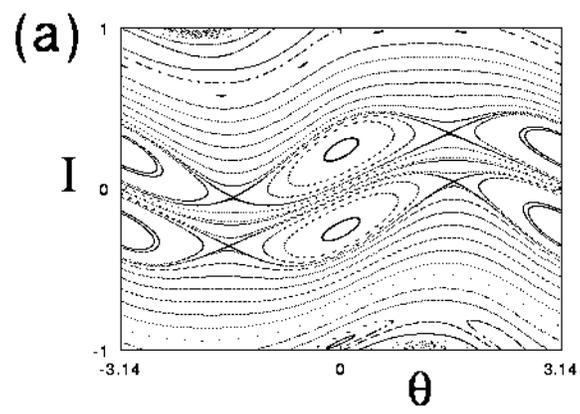
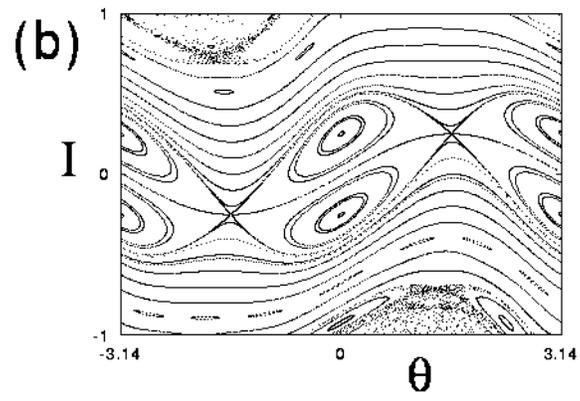
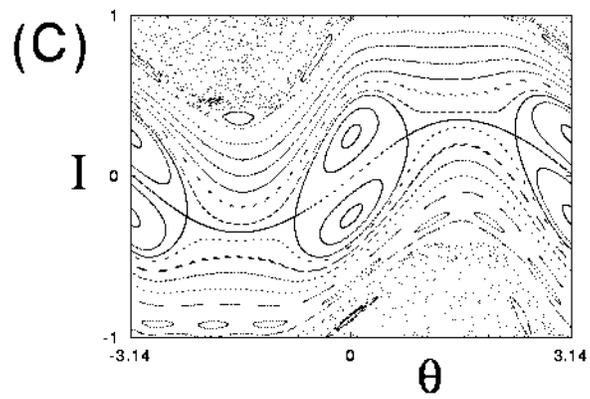
{\bf Fig.2.}
\begin{minipage}[t]{145mm}
The merging process of the twin-chains with period-two for various
values of $K$ at $\mu=-0.49$.  (a)$~K$=0.4(below
threshold). ~(b)$~K$=0.50132565(reconnection
threshold). ~(c)$~K$=0.7(above threshold).
\end{minipage}
\vskip 3mm
\noindent
{\bf Fig.3.}
\begin{minipage}[t]{145mm}
The breakup diagram of shearless KAM curves and the reconnection
thresholds for the twin-chains with rational rotation numbers
$P/Q$. The critical point $(\mu_c=-0.313951, K_c=1.54156)$ obtained by
del-Castillo-Negrete,~Greene and Morrison is indicated by an arrow.
\end{minipage}
\vskip 3mm
\noindent
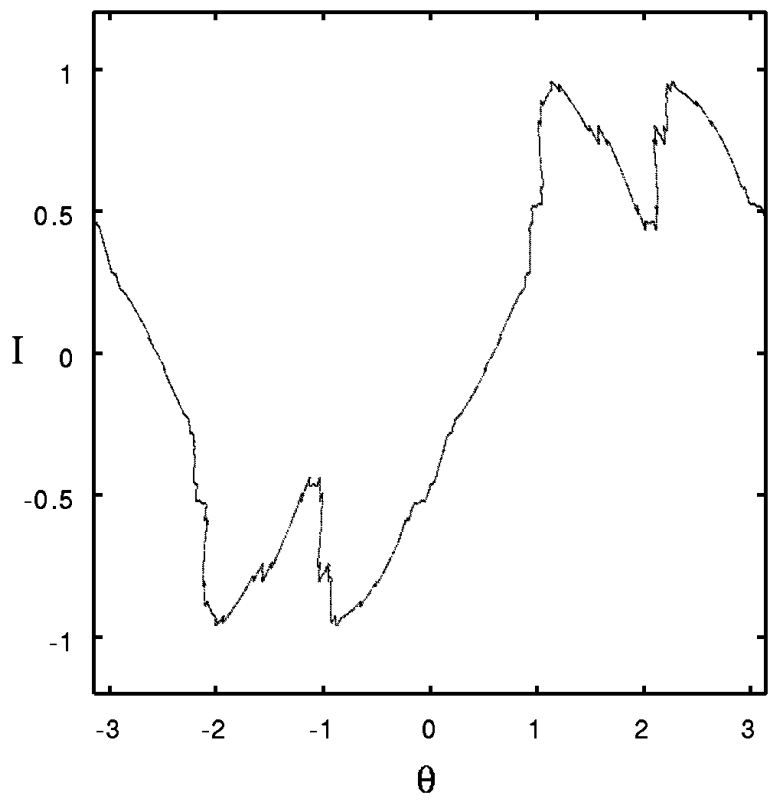
{\bf Fig.4.}
\begin{minipage}[t]{145mm}
The shearless KAM curve for $(\mu_c,K_c)$.
\end{minipage}
\vskip 3mm
\noindent
{\bf Fig.5.}
\begin{minipage}[t]{145mm}
A magnification of the breakup diagram around $(\mu_c,K_c)$.
\end{minipage}
\end{document}